\documentclass[pra,aps,superscriptaddress,amssymb,amsmath,reprint,noeprint,floatfix]{revtex4-2}
\usepackage{natbib}
\usepackage{braket}
\usepackage{bm}
\usepackage[normalem]{ulem}
\usepackage{siunitx}
\usepackage{soul,xcolor}

\usepackage{graphicx}
\usepackage[colorlinks=true,allcolors=blue]{hyperref}%
\usepackage{xcolor}
\hypersetup{bookmarksnumbered}

\newcommand{\ve}[1]{\mathbf{#1}}

\newcommand{\epsilonr}{\varepsilon_{r}}
\newcommand{\Q}{\Omega}
\newcommand{\Lamb}{\Lambda}

\begin{document}
\setstcolor{red}

\title{\textcolor{black}{Strong optical nonreciprocity in a photonic crystal composed of spinning cylinders}}
\date{\today}

\author{Hengzhi Li}
\affiliation{Department of Physics, City University of Hong Kong, Tat Chee Avenue, Kowloon, Hong Kong, China}
\author{Wanyue Xiao}
\affiliation{Department of Physics, City University of Hong Kong, Tat Chee Avenue, Kowloon, Hong Kong, China}
\author{Junho Jung}
\affiliation{Department of Physics, City University of Hong Kong, Tat Chee Avenue, Kowloon, Hong Kong, China}
\author{Hao Pan}
\affiliation{Department of Physics, City University of Hong Kong, Tat Chee Avenue, Kowloon, Hong Kong, China}
\author{Shubo Wang}\email{shubwang@cityu.edu.hk}
\affiliation{Department of Physics, City University of Hong Kong, Tat Chee Avenue, Kowloon, Hong Kong, China}

\begin{abstract}
 Moving media break time-reversal symmetry and exhibit intriguing optical nonreciprocity. This nonreciprocity is usually weak due to the much lower moving speed of media relative to the speed of light. We demonstrate that strong optical nonreciprocity can emerge in a two-dimensional photonic crystal composed of spinning dielectric cylinders. The photonic crystal supports two types of chiral modes at the Brillouin zone center: hybridized multipole modes and symmetry-protected bound states in the continuum (BICs), both of which carry intrinsic spin angular momentum. For finite wavevectors near the zone center, the BICs transform into quasi-bound states in the continuum (QBICs). Under oblique incidence of circularly polarized plane waves, the photonic crystal exhibits nonreciprocal transmission and absorption that are significantly enhanced at the frequencies of these hybridized multipole modes and QBICs.  Furthermore, the high quality factors of the QBICs enable sharp transitions in nonreciprocity.  Our work uncovers strong chiral light-matter interactions in periodic moving structures, with potential applications in nonreciprocal light manipulation. The mechanism may also be generalized to other classical wave systems, such as phononic crystals. 
\end{abstract}
\maketitle

\section{\label{sec: I. Introduction}Introduction}

Lorentz reciprocity is a natural property of electromagnetic systems arising from the time-reversal invariance of Maxwell’s equations \cite{caloz2018electromagnetic}. By breaking the time-reversal symmetry, nonreciprocal systems exhibit intriguing optical properties with rich applications, such as optical isolation and circulation \cite{lira2012electrically,estep2014magnetic}. Electromagnetic nonreciprocity can be realized by exploiting three major physical mechanisms: magneto-optical effects \cite{palatnik2021possible,dotsch2005applications,pan2024nonreciprocal}, optical nonlinearity \cite{shi2015limitations,khanikaev2015nonlinear,peng2014parity,guo2022nonreciprocal}, and temporal modulations \cite{yu2009complete,guo2019nonreciprocal,estep2014magnetic,taravati2017nonreciprocal}. Magneto-optical effects introduce asymmetry in the permittivity or permeability of materials under an external magnetic field bias. Optical nonlinearity relies on nonlinear materials with a refractive index that varies with local electric field intensity. Temporal modulations render a media temporally inhomogeneous, thereby breaking the time-reversal symmetry. Recently, spinning motion has emerged as an alternative mechanism for realizing electromagnetic nonreciprocity \cite{Huang2018,jiao2020nonreciprocal,Yang2022}. Spinning motion transforms normal isotropic media into effectively bianisotropic media of the Tellegen type \cite{li2025optical}, which inherently breaks time-reversal symmetry and exhibits nonreciprocal properties. For instance, a single spinning cylinder can give rise to optical isolation of guided waves due to the unidirectional coupling protected by the transverse spin-orbit interaction \cite{Shi2021,Yang2022}. Furthermore, a one-dimensional lattice of spinning cylinders can enable nonreciprocal wavefront manipulation, giving rise to one-way deflection of incident plane waves \cite{yang2025nonreciprocal}. \textcolor{black}{However, optical nonreciprocity induced by spinning motion is usually weak due to the much lower spinning speed compared to the speed of light. Thus, exploring mechanisms to enhance the nonreciprocity is essential for practical applications of spinning systems.}


In recent years, bound states in the continuum (BICs) have attracted considerable attention for their crucial role in strong wave-matter interaction phenomena, such as lasing \cite{kodigala2017lasing,wang2025inherent} , electromagnetically induced transparency \cite{koshelev2018asymmetric,kutuzova2025polarization}, and circular dichroism \cite{gorkunov2020metasurfaces,shi2022planar}. BICs can emerge in various optical structures, including optical gratings \cite{marinica2008bound}, photonic crystals \cite{hsu2013observation}, and metasurfaces \cite{koshelev2018asymmetric}. The emergence of BICs in periodic structures is closely related to polarization singularities \cite{zhen2014topological,cui2025ultracompact,nye1981motion}, which exhibit rich topological properties and intriguing geometric phase phenomena \cite{fu2024near,cheng2025riemann,wang2020generating}. Ideal BICs cannot be excited by external excitations due to the complete suppression of radiation. By introducing perturbations to break the symmetry of BIC systems, such as via oblique incidence or  geometric deformations, the quasi-bound states in the continuum (QBICs) can be obtained, which exhibit strongly localized fields and high Q-factors \cite{doiron2022realizing,chen2019strong,li2024nonreciprocal}. Consequently, QBICs have been employed to enhance electromagnetic nonreciprocity in artificial structures \cite{manez2024extreme,cai2024nonreciprocal,li2025nonreciprocal}. However, these systems rely on the conventional mechanisms (i.e., magneto-optical effects, nonlinearity, or temporal modulations) to achieve nonreciprocity. The effect of QBICs on the spinning-motion-induced nonreciprocity has remained elusive.


In this work, we investigate electromagnetic nonreciprocity in a two-dimensional (2D) photonic crystal composed of finite-sized spinning dielectric cylinders. \textcolor{black}{We show that the photonic crystal supports chiral QBICs and hybridized multipole modes in the vicinity of Brillouin zone center. Under oblique incidence of plane waves that break the symmetry, both the QBICs and hybridized multipole modes can give rise to strongly enhanced nonreciprocity in transmission, absorption, and circular dichroism (CD). Compared with the hybridized multipole modes, the QBICs can enable sharp transitions in nonreciprocity due to their high Q-factors.} The results elucidate the interplay between BICs/QBICs and relativistic effects in determining the nonreciprocal properties of moving structures, providing important insights into the different characteristics of moving media and time-varying media.

The article is organized as follows. In Sec. II, we introduce the methodology for simulating the photonic crystal composed of spinning cylinders. We also introduce the semi-analytical multipole expansion method used to understand the properties of the eigenmodes in the spinning cylinders. In Sec. III, we first discuss the properties of the band structures and eigenmodes of the photonic crystal. Then, we discuss the nonreciprocal properties of the photonic crystal enhanced by the hybridized multipole modes and QBICs. The conclusion is drawn in Sec. IV.

\section{\label{sec: II. MST,SourceRep}	METHODOLOGY }

Objects in uniform motion are effectively bianisotropic due to special relativistic effect \cite{minkowski1908grundgleichungen,sommerfeld2013electrodynamics}. For an axially symmetric body in spinning motion about its axis, like a sphere or a cylinder, its bianisotropic electromagnetic properties can be described by the following constitutive relations \cite{Ridgely1998,Ridgely1999}
\begin{equation}
\begin{aligned}
   & \mathbf{D}+\mathbf{v} \times \frac{\mathbf{H}}{c^{2}}=\varepsilon(\mathbf{E}+\mathbf{v} \times \mathbf{B}),\\
   & \mathbf{B}+\mathbf{E} \times \frac{\mathbf{v}}{c^{2}}=\mu(\mathbf{H}+\mathbf{D} \times \mathbf{v}),
\label{eqn:1}
\end{aligned}
\end{equation}
where $\varepsilon$ and $\mu$ are the absolute permittivity and permeability of the material at rest; $\mathbf{v}=\mathbf{v}(\mathbf{r})$ is the linear velocity of a point on the object. \textcolor{black}{SI units are employed throughout the paper.} The above constitutive relations can be formulated into a matrix form:
\begin{equation}
\left[\begin{array}{l}
\mathbf{D} \\
\mathbf{B}
\end{array}\right]=\left[\begin{array}{cc}
\bar{\varepsilon} & \bar{\chi}_{\mathrm{em}} \\
\bar{\chi}_{\mathrm{me}} & \bar{\mu}
\end{array}\right]\left[\begin{array}{l}
\mathbf{E} \\
\mathbf{H}
\end{array}\right],
\label{eqn:2}
\end{equation}
where $\bar{\varepsilon}$, $\bar{\mu}$, $\bar{\chi}_{\mathrm{me}}$, and $\bar{\chi}_{\mathrm{em}}$ are rank two tensors that characterize the effective material properties of the spinning sphere. They have the following elements in spherical coordinates system: 
\textcolor{black}{
\begin{equation}
\begin{aligned}
&\bar{\varepsilon}
=\frac{1}{\beta}\left[\begin{array}{ccc}
\alpha \varepsilon & 0 & 0 \\
0 & \alpha\varepsilon & 0\\
0 & 0 & \beta \varepsilon
\end{array}\right],
\bar{\chi}_{\mathrm{em}}
=\frac{1}{\beta}\left[\begin{array}{ccc}
0 &  \gamma  & 0 \\
-\gamma  & 0 & 0\\
0 & 0 & 0
\end{array}\right], \\
&\bar{\chi}_{\mathrm{me}}
=\frac{1}{\beta}\left[\begin{array}{ccc}
0 & -\gamma  & 0 \\
\gamma  & 0 & 0\\
0 & 0 & 0
\end{array}\right],
\bar{\mu}
=\frac{1}{\beta}\left[\begin{array}{ccc}
\alpha \mu & 0 & 0 \\
0 & \alpha \mu & 0\\
0 & 0 & \beta \mu
\end{array}\right].
\end{aligned}
\label{eqn:3}
\end{equation}
}
\noindent Here, $\alpha=c^{2}-v^{2}$, $\beta=c^{2}\left(1-\epsilon \mu v^{2}\right)$, $\gamma=(1-\varepsilon\mu c^{2})v$. $c$ is the speed of light in vacuum. $v=\varOmega r \text{sin}\theta$ with $\varOmega$ and $r$ being the angular velocity and radial distance. \textcolor{black} The bianisotropy is characterized by the matrices $\bar{\chi}_{\mathrm{em}}=\bar{\chi}_{\mathrm{me}}^\text{T}$ , indicating that the spinning objects can be considered as made of an effective Tellegen-type material. In addition, the spinning motion can give rise to an effective gauge field acting on electromagnetic waves, modulating the waves’ phase in the planes perpendicular to the spinning axis \cite{shi2019gauge}. The above constitutive relations can be implemented in COMSOL (a finite-element-method package) to simulate the electromagnetic properties of spinning objects.

To investigate the eigenmodes of the photonic crystal, we apply multipole expansions in source representation to decompose the eigenmodes into electromagnetic multipoles. The multipoles can be evaluated using the currents excited inside the unit cells for the lowest few orders, as \cite{Alaee2018,Mobini2018}
\begin{equation}
\begin{aligned}
p_{i}=&-\frac{1}{i \omega} \int d\tau J_{i} j_{0}(k r)\\&+
\frac{ik^{2}}{2\omega} \left\{\int d\tau\left[3(\mathbf{r} \cdot \mathbf{J}) r_{i}
-r^{2} J_{i}\right] \frac{j_{2}(k r)}{(k r)^{2}} \right\}
\end{aligned}
\label{eqn:11}
\end{equation}
\begin{equation}
m_{i}=\frac{3}{2} \int d\tau(\mathbf{r} \times \mathbf{J})_{i} \frac{j_{1}(k r)}{k r},
\label{eqn:12}
\end{equation}
\begin{equation}
\begin{aligned}
Q_{i j}^{e} = & -\frac{3}{i \omega} \left\{ \int d \tau [3(r \odot J) - 2(\mathbf{r} \cdot \mathbf{J}) \delta_{i j}] \frac{j_{1}(k r)}{k r} \right. \\
& + 2 k^{2} \int d \tau [5 r_{i} r_{j}(\mathbf{r} \cdot \mathbf{J}) - (r \odot J) r^{2} \\
& \left. - r^{2}(\mathbf{r} \cdot \mathbf{J}) \delta_{i j}] \frac{j_{3}(k r)}{(k r)^{3}} \right\},
\end{aligned}
\end{equation}
\begin{equation}
Q_{ij}^{m}=15 \int d\tau \ [r_{i}(\mathbf{r} \times \mathbf{J})_{j}+r_{j}(\mathbf{r} \times \mathbf{J})_{i} ] \frac{j_{2}(k r)}{(k r)^{2}},
\label{eqn:14}
\end{equation}
\noindent where $J_i$ are the Cartesian components of the induced electric current density $\mathbf{J}$, $k$ is the wave number of the incident plane wave, $r\odot J=r_{j} J_{i}+r_{i} J_{j}$, and $j_{1,2,3} (kr)$ are the spherical Bessel functions. $p_i$, $m_i$, $Q_{ij}^e$, $Q_{ij}^m$ are the Cartesian components of the electric dipole, magnetic dipole, electric quadrupole, and magnetic quadrupole, respectively.

\section{\label{sec: III. MST,SourceRep} RESULTS AND DISCUSSION }

\subsection{\label{sec: A} Band structures and eigenmodes}
We consider the photonic crystal shown in Fig. 1(a), which is composed of silicon cylinders ($\epsilonr = 11.9$) arranged into a square lattice in air with lattice constant $D = 600$\,nm. The cylinder in each unit cell has the radius $a = 200$\,nm and the height $l = 390$\,nm, and it spins around its center axis with an angular velocity $\ve{\Q} = \Q \hat{z}$. We define the normalized rotation speed $\Lamb = \Q a / c$ to be used in the following discussions. All numerical simulation results in this article are obtained by using COMSOL.

\begin{figure*}[htb]
    \centering
    \includegraphics[width=0.9\linewidth] {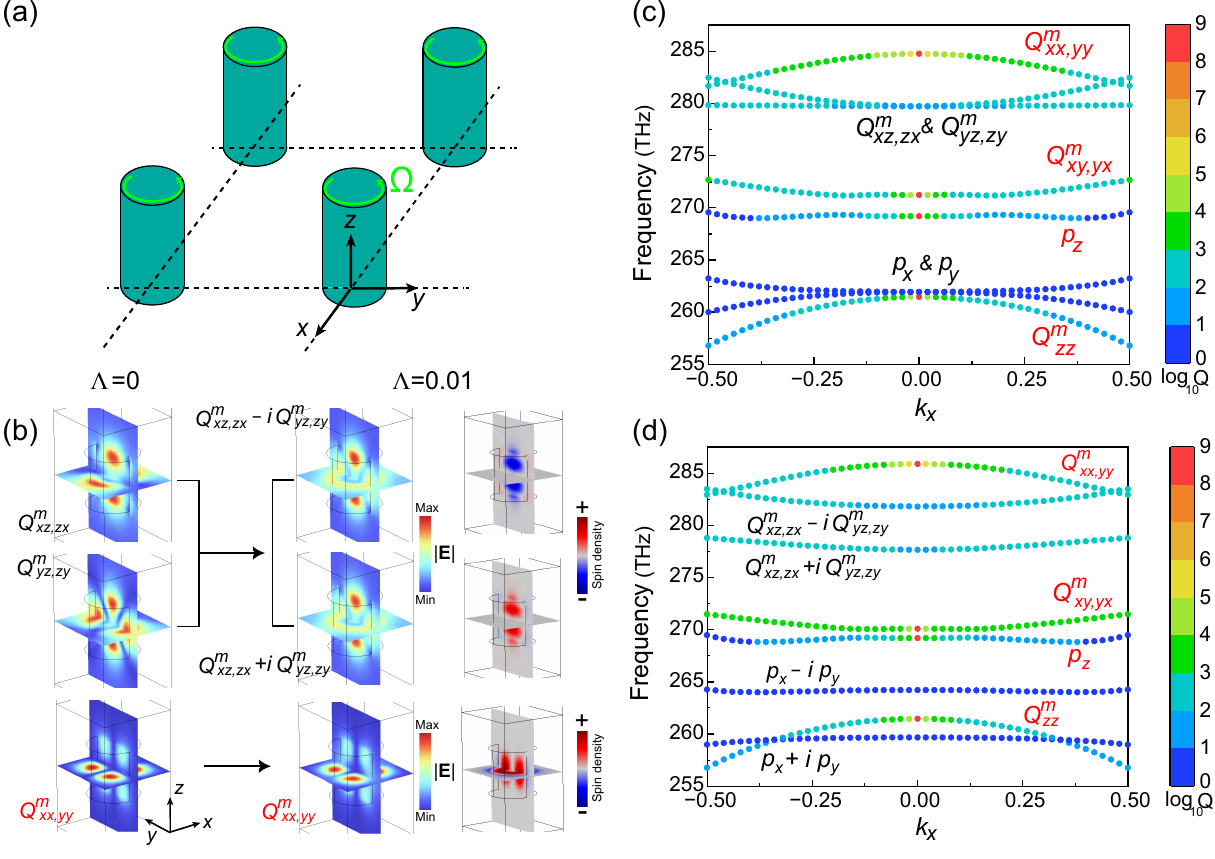}
    \caption{(a) Schematic of the photonic crystal composed of spinning dielectric cylinders arranged in a square lattice with period D = 600 nm. The cylinder in each unit cell has the radius $a$ = 200 nm and the height $l$ = 390 nm, and its spinning angular velocity is $\ve{\Q}=\varOmega \hat{z}$. (b) Electric field norm and spin density of the eigenmodes at $\mathrm{\Gamma}$ point of the 6th, 7th and 8th bands (counted from bottom to top) in (c). (c) Band structure of the photonic crystal with rotation speed $\Lambda$=0. The color of the bands denotes the quality factor of the eigenmodes. The bands are labeled by their dominant multipole components. (d) Band structure of the photonic crystal with rotation speed $\Lambda$=0.01.}
    \label{fig:1}
\end{figure*}

We compute the band structures of the photonic crystal with $\Lamb = 0$ and $\Lamb = 0.01$. The results are shown in Figs. 1(c) and 1(d), respectively, which enables a direct comparison of the band structures of the stationary and spinning systems. The color of the bands denotes the Q-factor of the corresponding eigenmodes. We focus on the bands within the frequency range of $[200, 300]$\,THz and apply \textcolor{black}{Eqs. (4)-(7)} to calculate the dominating multipole components of the eigenmodes, which are labeled near the bands in Fig. 1(c). We notice that four red points at the zone center have $Q \rightarrow \infty$ (the maximum value of the colorbar is capped at $10^9$), corresponding to four BICs. The emergence of these BICs is attributed to the symmetry of the system. The lowest two BICs (counted from bottom to top)  are dominated by the magnetic quadrupole $Q_{zz}^{m}$ and electric dipole $p_{z}$, respectively, whose fields satisfy the cylindrical symmetry with respect to $z$-axis and cannot couple to the radiation channels with $k_{x} = k_{y} = 0$. The third BIC is dominated by the magnetic quadrupole components $Q_{xy}^{m}$ and $Q_{yx}^{m}$ with equal amplitudes, which satisfy the $C_{2}$ rotational symmetry with respect to $z$-axis and thereby cannot radiate due to the mismatch with the symmetry of free-space channels. The fourth BIC is dominated by the magnetic quadrupole components $Q_{xx}^{m}$ and $Q_{yy}^{m}$ with $Q_{xx}^{m} = - Q_{yy}^{m}$, whose fields also exhibit a $C_{2}$ rotational symmetry and cannot be coupled to free-space channels. Consequently, the emergence of all the four BICs at the $\Gamma$ point is protected by the symmetries that forbid their radiation to free space.

In addition to the BICs, this stationary photonic crystal also supports non-BIC eigenmodes at the $\Gamma$ point, which exhibit degeneracies due to the $C_{4}$ symmetry \textcolor{black}{and time-reversal symmetry (or the $\sigma_v$ symmetry)} of the system. Specifically, the 2$^{\text{nd}}$ and 3$^{\text{rd}}$ lowest bands in Fig. 1(c) are degenerate at the zone center, corresponding to the orthogonal electric dipole modes $p_{x}$ and $p_{y}$. The 6$^{\text{th}}$ and 7$^{\text{th}}$ bands are also degenerate at the zone center, corresponding to two orthogonal magnetic quadrupole modes with the components $Q_{xz,zx}^{m}$ and $Q_{yz,zy}^{m}$, respectively. We note that the considered photonic crystal does not support symmetry-protected \emph{degenerate} BICs, which would require at least a $C_{6}$ symmetry \textcolor{black}{and time-reversal symmetry} of the photonic crystal \cite{cheng2025riemann}.


Figure 1(d) shows the band structure of the photonic crystal with normalized rotation speed $\Lamb = 0.01$. We notice that the spinning motion shifts the eigenfrequencies of the BICs dominated by $Q_{xy,yx}^{m}$ and $Q_{xx,yy}^{m}$, while the eigenfrequencies of the BICs dominated by $Q_{zz}^{m}$ and $p_{z}$ remain almost unchanged. This is because the fields of $Q_{zz}^{m}$ and $p_{z}$ are linearly polarized along $z$ direction and cylindrically symmetric (without azimuthal phase variation), which are nearly independent of the effective gauge field induced by spinning motion \cite{shi2019gauge,Yang2022}. In contrast, the fields of $Q_{xy,yx}^{m}$ and $Q_{xx,yy}^{m}$ are $C_{2}$ symmetric and have an azimuthal wavevector component, which are strongly modified by the effective gauge field. Notably, under the spinning motion of the cylinders, the BICs dominated by $Q_{xy,yx}^{m}$ and $Q_{xx,yy}^{m}$ turn into chiral BICs carrying spin angular momentum. This is shown in the last column of Fig. 1(b), where the BIC dominated by $Q_{xx,yy}^{m}$ has the eigenfield rotating around $z$-axis and carrying spin angular momentum in $+z$ direction. Similar properties also exist in the BIC dominated by $Q_{xy,yx}^{m}$. We note that the spin density has opposite signs for the internal field and the near field outside the cylinder; the internal spin determines the spin of the radiation fields.


In Fig. 1(d), the spinning motion lifts the degeneracy of the electric dipole modes $p_{x}$ and $p_{y}$, as well as the degeneracy of the magnetic quadrupole modes $Q_{xz,zx}^{m}$ and $Q_{yz,zy}^{m}$ at the $\Gamma$ point. This is attributed to the hybridization of the orthogonal multipole components induced by the effective bianisotropy. Taking the case of the magnetic quadrupole modes as an example, the spinning-induced bianisotropy mixes the orthogonal components $Q_{xz,zx}^{m}$ and $Q_{yz,zy}^{m}$, giving rise to a pair of nondegenerate chiral modes $Q_{xz,zx}^{m} + iQ_{yz,zy}^{m}$ and $Q_{xz,zx}^{m} - iQ_{yz,zy}^{m}$, which carry opposite spin angular momentum, as shown in Fig. 1(b). This degeneracy breaking can also be understood as due to the effective gauge field induced by the spinning motion. Similar properties also exist in the electric dipole modes $p_{x}$ and $p_{y}$.



\subsection{\label{sec: B} Nonreciprocity enhanced by chiral hybridized multipole modes}
We consider the photonic crystal with the rotation speed $\Lamb = 0.01$, under the incidence of a right-handed circularly polarized (RCP) plane wave $\mathbf{E} = (\widehat{x} + i\widehat{y})E_{0}e^{- ikz - i\omega t}$, as shown in Fig. 2(a). Figure 2(b) shows the numerically simulated transmission spectra under backward ($+z$) and forward ($-z$) incidences, which are denoted as $T_{\mathrm{f}}$ (solid blue line) and $T_{\mathrm{b}}$ (solid red line), respectively. For comparison, we also show the transmission spectra of the stationary photonic crystal with $\Lamb = 0$, corresponding to the solid black line. In this case, the system maintains the $C_{2}$ symmetry, and the BICs cannot be excited by the incident plane wave. In contrast, the chiral hybridized multipole modes can be excited, manifesting as dips in the transmission spectra of the system. For the stationary system, we notice that two dips appear at about 250 THz and 280 THz, respectively. The first dip at 250 THz is attributed to the excitation of the degenerate dipole mode $p_{x}$ and $p_{y}$ in Fig. 1(b). The second dip at 280 THz is attributed to the excitation of the degenerate magnetic quadrupole modes $Q_{xz,zx}^{m}$ and $Q_{yz,zy}^{m}$ in Fig. 1(b). For the spinning system, we notice that the transmission dip at 280 THz is shifted towards opposite directions under the forward and backward incidences. This is because that the forward and backward incident RCP plane waves carry opposite spin angular momentum. Consequently, they excite the opposite chiral quadrupoles $Q_{xz,zx}^{m} + iQ_{yz,zy}^{m}$ and $Q_{xz,zx}^{m} - iQ_{yz,zy}^{m}$ in Fig. 1(b), which have different resonance frequencies due to the Sagnac effect \cite{sagnac1913ether}. The different locations of the transmission dips of $T_{\mathrm{f}}$ and $T_{\mathrm{b}}$ lead to a large transmission contrast $T_{\mathrm{f}} - T_{\mathrm{b}}$, i.e., nonreciprocal isolation ratio, as denoted by the green line in Fig. 2(b), showing two local extrema near the quadrupole resonance frequencies. The maximum isolation ratio reaches nearly 100\%, demonstrating strong nonreciprocity induced by the chiral hybridized multipole modes in the photonic crystal.

\begin{figure}[tb]
    \centering
    \includegraphics[width=\linewidth]{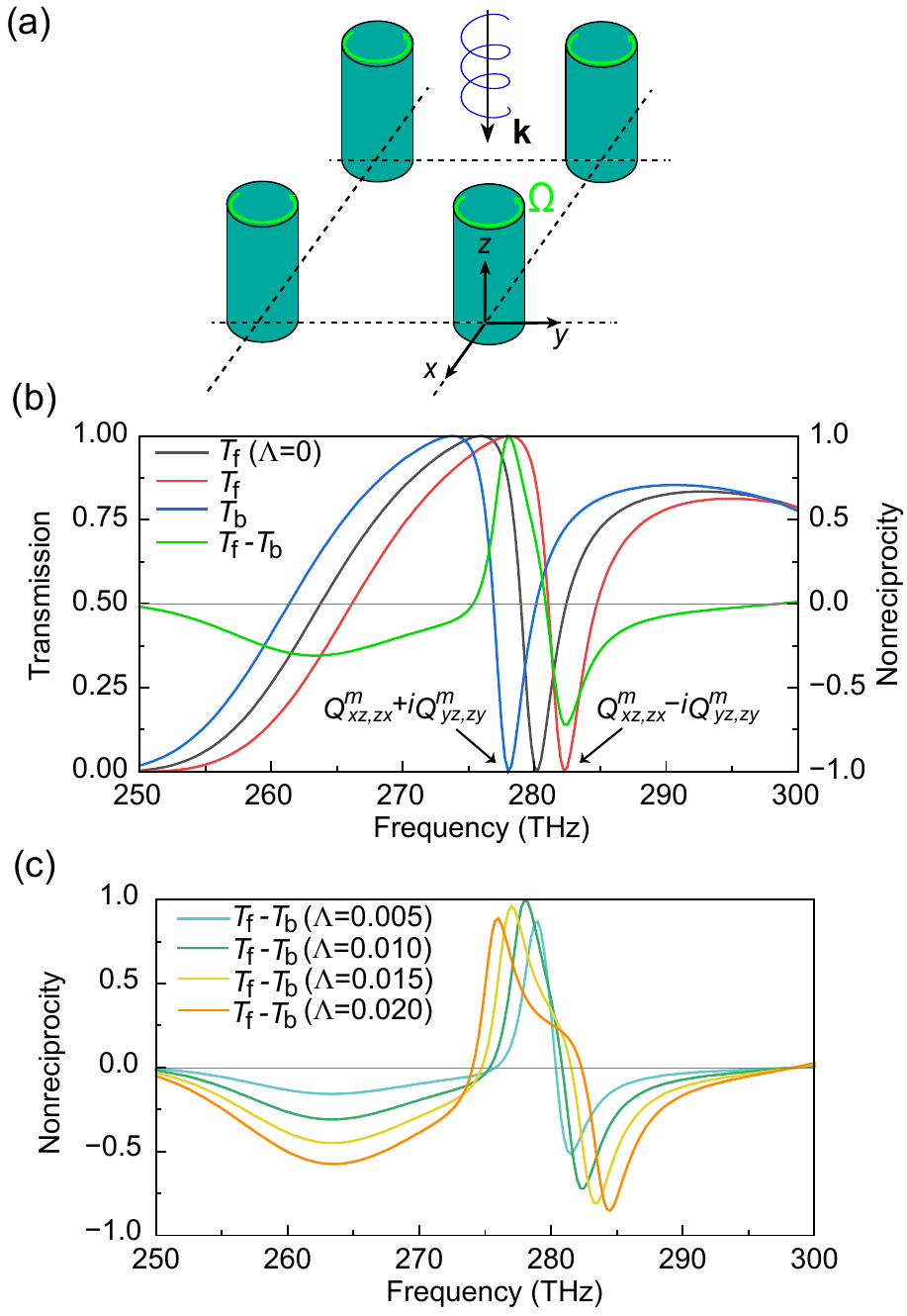}
    \caption{(a) Schematic of the photonic crystal under the forward incidence of a circularly polarized plane wave. (b) Transmission spectra of the photonic crystal under forward ($T_{\mathrm{f}}$) and backward ($T_{\mathrm{b}}$) incidences. We set the normalized spinning speed $\Lambda=0.01$. The black solid line denotes the result of the stationary photonic crystal. (c) Transmission contrast $T_{\mathrm{f}}-T_{\mathrm{b}}$ as a function of the normalized spinning speed $\Lambda$. }
    \label{fig:2}
\end{figure}

To understand the dependence of nonreciprocity on the spinning motion of the cylinders, we calculate the transmission contrast $T_{\mathrm{f}} - T_{\mathrm{b}}$ under different spinning speeds $\Lamb$. The results are summarized in Fig. 2(c). We notice that the transmission contrast changes sign near the quadrupole resonance frequency 280 THz due to the selective excitation of the chiral quadrupoles $Q_{xz,zx}^{m} \pm iQ_{yz,zy}^{m}$ by the opposite incident waves. In addition, the absolute nonreciprocity, i.e., $|T_{\mathrm{f}} - T_{\mathrm{b}}|$, generally increases with spinning speed $\Lamb$. Notably, as $\Lamb$ increases, the peak value of $T_{\mathrm{f}} - T_{\mathrm{b}}$ first approaches 100\% and then decreases, indicating an optimal rotation speed for achieving optical isolation. These results demonstrate tunable nonreciprocity of the photonic crystal by adjusting the spinning speed. Similar nonreciprocal properties also exist under the incidence of left-handed circularly polarized (LCP) plane waves.

\subsection{\label{sec: C} Nonreciprocity enhanced by chiral QBICs}

While the BICs at the $\Gamma$ point cannot be excited by free-space plane waves, the QBICs at a finite $\mathbf{k}$ can be excited by incident plane waves. These QBICs can also influence the nonreciprocal properties of the photonic crystal. To understand this influence, we consider an RCP plane wave $\mathbf{E} = E_{0}e^{ik\left( \sin\theta x - \cos\theta z \right) - i\omega t}$ obliquely incident on the structure with the incident angle $\theta = 5^{\circ}$, as shown in Fig. 3(a). Figure 3(b) shows the numerically simulated transmission spectra under forward ($\theta = 5^{\circ}$) and backward ($\theta = 185^{\circ}$) incidences as well as the transmission contrast $T_{\mathrm{f}}-T_{\mathrm{b}}$. The locations of the QBICs are marked by the black arrows. Compared to the transmission spectra in Fig. 2(b), we notice that the QBICs induce additional Fano-type resonances with dramatic changes of transmission in narrow frequency ranges. The influence of the QBICs also manifests in the transmission contrast denoted by the green line. We notice that the QBICs can either increase or reduce the transmission contrast. Specifically, the first three QBICs (labeled as $Q_{zz}^{m}$, $p_{z}$, and $Q_{xy,yx}^{m}$) lead to near-zero transmission contrast. This is because these QBICs contribute to a positive contrast $T_{\mathrm{f}}-T_{\mathrm{b}}$, which cancels out the broad background of negative contrast resulting from the hybridization of $p_{x}$ and $p_{y}$. Interestingly, the fourth QBIC (labeled as $Q_{xx,yy}^{m}$) enables a sharp transition of the transmission contrast from about $-1$ (nearly-perfect isolation) to about zero (nearly-symmetric transmission) at 286 THz. Compared to the cases of the chiral hybridized multipole modes in Sec. IIIB, the QBICs here enable sharp modulations of the nonreciprocity due to their high Q-factors.

To understand how the nonreciprocity depends on the incident angle, we computed the transmission contrast $T_{\mathrm{f}}-T_{\mathrm{b}}$ for different incident angles $\theta = 5^{\circ},10^{\circ},$ and $15^{\circ}$, as shown in Fig. 3(c). As seen, the QBICs-induced peaks and dips are shifted in the spectrum, and their Q-factors reduce as $\theta$ increases. This is consistent with the eigen properties of the system shown in Fig. 1(d), where the Q-factors of the modes decrease as $k_{x}$ increases. In addition, it is interesting to notice that the nonreciprocity (i.e., $|T_{\mathrm{f}}-T_{\mathrm{b}}|$) induced by the QBIC with dominating $Q_{xy,yx}^{m}$ gradually increases with $\theta$, despite the reduction of its Q-factor, which is attributed to the interference between $Q_{xx,yy}^{m}$ and the chiral hybridized multipole $Q_{xz,zx}^{m} + iQ_{yz,zy}^{m}$.

\begin{figure}[tb]
    \centering
    \includegraphics[width=1\linewidth]{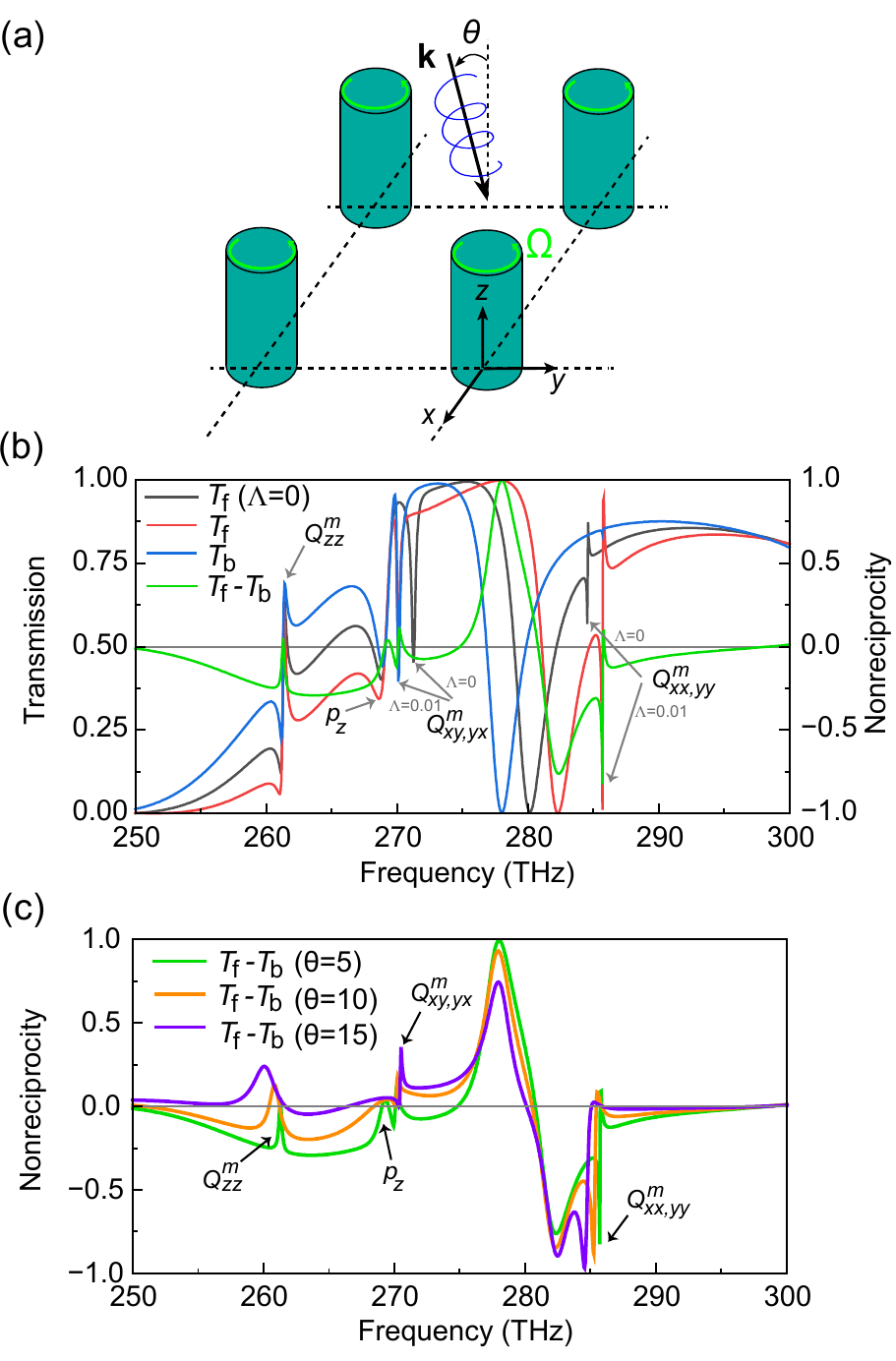}
    \caption{(a) Schematic of the photonic crystal under the oblique incidence of a circularly polarized plane wave with incident angle $\theta$. (b) Transmission spectra of the photonic crystal under forward ($T_{\mathrm{f}}$) and backward ($T_{\mathrm{b}}$) incidences. We set the normalized spinning speed $\Lambda$=0.01. The black solid line denotes the result
of the stationary photonic crystal. (c) Transmission contrast $T_{\mathrm{f}}-T_{\mathrm{b}}$ for different incident angles $\theta=5^\circ,10^\circ, 15^\circ$.}
    \label{fig:3}
\end{figure}

Finally, we consider the photonic crystal composed of spinning lossy cylinders with relative permittivity $\varepsilon_{r} = 11.9 + 0.01i$. In this case, the structure can give rise to both nonreciprocity and circular dichroism (CD). Thus, it is essential to understand the interplay between CD and nonreciprocity. Figure 4(a) shows the schematics for demonstrating CD and nonreciprocity in this system. The CD refers to the differential absorption of the photonic crystal under the forward incidence of RCP and LCP lights, corresponding to system A and system B1, respectively. The nonreciprocity refers to the differential transmission of the photonic crystal under the forward and backward incidences of RCP light, corresponding to system A and system B2, respectively. Since systems B1 and B2 can be transformed into each other through a mirror operation with respect to $xoz$-plane and a $180^{\circ}$ rotation around $y$-axis, they must have the same absorption spectrum and the same transmission spectrum. Consequently, the CD is equivalent to the nonreciprocity in the considered photonic crystal system. 

\begin{figure}[tb]
    \centering
    \includegraphics[width=1\linewidth]{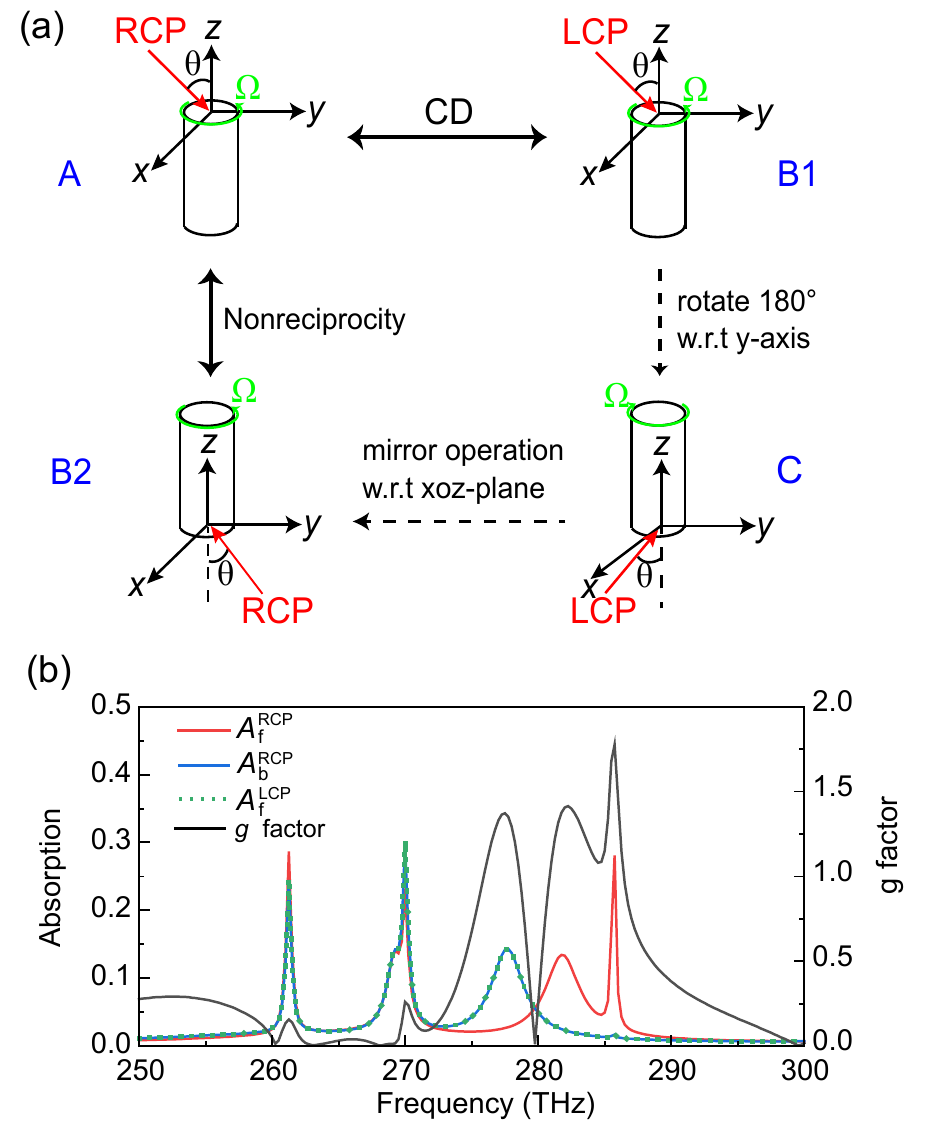}
    \caption{(a) Schematics for demonstrating CD and nonreciprocity in the photonic crystal composed of spinning cylinders. Systems B1 and B2 can be transformed into each other through a mirror operation with respect to xoz-plane and a $180^\circ$ rotation with respect to y-axis. (b) Absorption spectra of the photonic crystal ($\Lambda=0.01$) under the incidence of circularly polarized lights. The black line denotes the absorption dissymmetry factor $g$.}
    \label{fig:4}
\end{figure}

\textcolor{black}{The equivalence between CD and nonreciprocity in the considered system} is verified by full-wave numerical simulations, as shown in Fig. 4(b), where the absorption under forward LCP incidence ($A_{\mathrm{f}}^{\mathrm{LCP}}$) equals the absorption under backward RCP incidence ($A_{\mathrm{b}}^{\mathrm{RCP}}$). The absorption under forward RCP incidence ($A_{\mathrm{f}}^{\mathrm{RCP}}$) is also shown in the figure. Clearly, the absorption is significantly enhanced by the chiral QBICs and chiral hybridized multipole modes. To characterize the CD, we calculate the absorption dissymmetry factor $g = 2| A_{\mathrm{f}}^{\mathrm{LCP}} - A_{\mathrm{f}}^{\mathrm{RCP}} |/|A_{\mathrm{f}}^{\mathrm{lcp}} + A_{\mathrm{f}}^{\mathrm{RCP}}|$, which is denoted by the solid black line. As we see, both the QBICs and hybridized multipole modes can enhance the dissymmetry factor $g$. Particularly, the CD contributed by the chiral QBIC $Q_{xx,yy}^{m}$ exhibits the maximum dissymmetry factor $g = 1.8$. In contrast, the CD contributed by the hybridized multipole modes exhibits a relatively smaller dissymmetry factor $g$ but a larger bandwidth due to the lower Q-factors of the modes.

\section{\label{sec: IV. MST,SourceRep} CONCLUSION}
In summary, we investigate the nonreciprocal properties of a photonic crystal composed of spinning dielectric cylinders. We find that this photonic crystal supports two types of modes in the vicinity of the Brillouin zone center, i.e., chiral QBICs and chiral hybridized multipole modes. These modes carry intrinsic spin angular momentum and, thus, have different amplitudes when excited by circularly polarized lights with opposite spin. We demonstrate that both the QBICs and hybridized multipole modes can significantly enhance optical nonreciprocity, leading to a large transmission contrast between forward and backward incidences. Compared to the hybridized multipole modes, the QBICs exhibit large Q-factors and can enable a sharp transition of nonreciprocity (from near-perfect isolation to near-perfect symmetric transmission) in a narrow frequency range, which may find applications in designing new optical switches. In addition, we demonstrate that the lossy photonic crystal can give rise to differential absorption of opposite circularly polarized lights, corresponding to the CD phenomenon, which is also significantly enhanced by the QBICs and hybridized multipole modes. 

Spinning-induced nonreciprocity can be realized with general materials and at broad frequencies, compared to the nonreciprocity originating from magneto-optical effects and optical nonlinearity. The results in this article contribute to the understanding of nonreciprocal properties of moving media and may find applications in the design of novel nonreciprocal devices. The mechanism may also be generalized to other classical wave systems, such as phononic crystals.

\section{\label{sec: VII. Acknowledgements}Acknowledgements}
The work described in this paper was supported by the National Natural Science Foundation of China (No. 12322416) and the Research Grants Council of the Hong Kong Special Administrative Region, China (Projects No. CityU 11308223 and No. AoE/P-502/20).\\

\textit{Data availability}---The data that support the findings of this article are not publicly available. The data are available from the authors upon reasonable request.

\bibliography{mendeley}
\end{document}